\documentclass[%
 reprint,
 superscriptaddress,
 amsmath,amssymb,
 aps,
 prb,
]{revtex4-2}

\usepackage{graphicx}
\usepackage{dcolumn}
\usepackage{bm}
\usepackage{siunitx}

\usepackage{color}

\newcommand{\magenta}[1]{\textcolor{black}{#1}}

\begin{document}

\preprint{APS/123-QED}

\title{Logarithmic Criticality in Transverse Thermoelectric Conductivity \\ of the Ferromagnetic Topological Semimetal CoMnSb}

\author{Hiroto Nakamura}
\email{hnakamura@issp.u-tokyo.ac.jp}
\affiliation{%
 Institute for Solid State Physics, University of Tokyo, Kashiwa 277-8581, Japan
}%

\author{Susumu Minami}%
\affiliation{%
 Department of Physics, The University of Tokyo, Hongo, Bunkyo-ku, Tokyo 113-0033, Japan
}%

\author{Takahiro Tomita}
\affiliation{%
 Institute for Solid State Physics, University of Tokyo, Kashiwa 277-8581, Japan
}%

\affiliation{%
 CREST, Japan Science and Technology Agency (JST), 4-1-8 Honcho Kawaguchi, Saitama 332-0012, Japan
}%

\author{Agustinus Agung Nugroho}
\affiliation{Faculty of Mathematics and Natural Sciences, Institut Teknologi Bandung, Jl. Ganesha 10, 40132 Bandung, Indonesia}

\author{Satoru Nakatsuji}
\thanks{To whom correspondence should be addressed:\\ satoru@phys.s.u-tokyo.ac.jp}
\affiliation{%
 Institute for Solid State Physics, University of Tokyo, Kashiwa 277-8581, Japan
}%

\affiliation{%
 Department of Physics, The University of Tokyo, Hongo, Bunkyo-ku, Tokyo 113-0033, Japan
}%
\affiliation{%
 CREST, Japan Science and Technology Agency (JST), 4-1-8 Honcho Kawaguchi, Saitama 332-0012, Japan
}%
\affiliation{%
 Trans-Scale Quantum Science Institute, University of Tokyo, 7-3-1 Hongo, Bunkyo-ku, Tokyo 113-0033, Japan
}%
\affiliation{%
 Institute for Quantum Matter and Department of Physics and Astronomy, Johns Hopkins University, Baltimore, Maryland 21218, USA
}%

\date{\today}

\begin{abstract}
We report the results of our experimental studies on the magnetic, transport and thermoelectric properties of the ferromagnetic metal CoMnSb.
Sizable anomalous Hall conductivity $\sigma_{yx}$ and transverse thermoelectric conductivity $\alpha_{yx}$ are found experimentally and comparable in size to the values estimated from density-functional theory. 
Our experiment further reveals that CoMnSb exhibits $-T\ln T$ critical behavior in $\alpha_{yx}(T)$, deviating from Fermi liquid behavior $\alpha_{yx}\sim T$ over a decade of temperature between \SI{10}{K} and \SI{400}{K}, similar to ferromagnetic Weyl and nodal-line semimetals. Our theoretical calculation for CoMnSb also predicts the $-T\ln T$ behavior when the Fermi energy locates near the Weyl nodes in momentum space.
\end{abstract}

\maketitle

Recent discovery of novel quantum phases due to nontrivial topology in electronic structure has sparked various trends in studies of correlated materials. In particular, significant progress has been made for developing various new materials and phenomena in the class of ``topological magnets," magnetic materials that have non-trivial topology in momentum space \cite{Armitage2018,Machida2010,Wan2011,Nakatsuji2015,Kuroda2017,Sakai2018,Liu2018,Ye2018,Belopolski2019,Liu2019,Sakai2020,Yin2020,Kang2020}. A prominent highlight is the striking observation of the large anomalous Hall effect (AHE) in the absence of magnetization in the non-collinear antiferromagnets Mn$_3X$ ($X$ = Sn, Ge) \cite{Nakatsuji2015,Kiyohara2016, Nayak2016, Chen2021}. Moreover, this has led to the detection of large anomalous Nernst effect (ANE) in an antiferromagnet and the discovery of the magnetic Weyl fermions in a correlated magnet \cite{Ikhlas2017,Kuroda2017}. The observation of large AHE and ANE without net magnetization in antiferromagnets confirms their intrinsic mechanism based on Berry curvature and the role of Weyl nodes as the sources and drains in momentum space.

Significant enhancement of the ANE is a hallmark of a topological magnet as it is the measure of the Berry curvature at the Fermi energy \cite{Xiao2006,Sharma2016,Ikhlas2017,Chen2021}. Thus, the ANE, which is normally of the order of \SI{0.1}{\micro V/K} in ferromagnets \cite{Chen2021}, could be largely enhanced if a ferromagnet has a nontrivial topological texture in momentum space, generating large Berry curvature near the Fermi energy. In fact, subsequent studies have further revealed that the topological ferromagnets such as ferromagnetic Weyl semimetal Co$_2$MnGa and nodal line semimetals Fe$_{3}X$ ($X$ = Ga, Al)  may lead to giant ANE that reaches $\sim \SI{6}{\mu V/K}$ at room temperature, breaking the conventional scaling law with magnetization and laying the foundation for the magnetic thermoelectric technology \cite{Sakai2018,Sakai2020,Xu2020,Simuda2020,Mizuguchi2019}. Another prominent feature of these topological ferromagnets is the logarithmic temperature dependence in the transverse thermoelectric conductivity, i.e., $\alpha_{yx} \sim -T\ln T$, in contrast with the Fermi liquid $T$-linear behavior, i.e., $\alpha_{yx} \sim T$ \cite{Sakai2018,Sakai2020}. If the topological electronic structure produces the extremely large net Berry curvature with a large density of states of the associated bands, the energy dependence of the transverse thermoelectric conductivity may have a logarithmic divergence near the chemical potential for such topological textures, thus producing \magenta{the breakdown of Fermi liquid behavior} \cite{Sakai2018,Sakai2020,Minami2020}. The above topological ferromagnets have provided specific examples of such topological textures, namely a flat band producing enhanced Berry curvature due to the quantum Lifshitz transition between type-I and type-II Weyl semimetals in Co$_2$MnGa \cite{Sakai2018} and the nodal Web structure in Fe$_{3}X$ \cite{Sakai2020}. 

In this paper, we report our experimental studies of a new candidate material of a topological ferromagnet, CoMnSb. In particular, we find a logarithmic temperature dependence of the transverse thermoelectric conductivity, $\alpha_{yx} \sim -T\ln T$. This deviation from the Fermi liquid behavior suggests that this material host a topological texture near the Fermi energy. In fact, our theoretical calculation indicates that CoMnSb is a ferromagnetic Weyl semimetal having two pairs of Weyl nodes near the Fermi energy.

CoMnSb is reported to have two types of crystal structure. One is the half-Heusler structure $XYZ$. Removing a half of $X$ element from the full-Heusler structure $X_2YZ$ renders a noncentrosymmetric structure. The Heusler compounds are known to harbor various topological phases including the magnetic Weyl semimetal state in Co$_2$MnGa \cite{Sakai2018} and the half-Heusler antiferromagnets $R$PtBi ($R=$ Gd, Nd) \cite{Hirschberger2016,Shekhar9140}. 
The ferromagnetic half-Heusler CoMnSb (reffered as ``CoMnSb-H'' in the following text) is also theoretically proposed to exhibit a large ANE compared to normal ferromagnets due to a large Berry curvature \cite{Minami2018}. However, no successful synthesis of this structure type has been reported to date \cite{Szytua1972}. The other structure type is the superstructure of the CoMnSb-H and possesses the centrosymmetry (Fig. 1(a)). Previous studies report a successful synthesis of single phase samples of this centrosymmetric structure type and find that the superstructure phase has a ferromagnetic ordering below the Curie temperature $T_{\mathrm{C}} \sim $ \SI{470}{K} \cite{Szytua1972,Ksenofontov2006, Nakamura2020}.

A single-crystalline sample of CoMnSb is grown by the method reported elsewhere. The compound is confirmed to crystallize in a 2 $\times$ 2 $\times$ 2 cubic superstructure of CoMnSb ($Fm\bar{3}m$) with a lattice parameter \mbox{$a$ = 11.835 \AA} \cite{Nakamura2020}. This superstructure and lattice parameter are used for the theoretical analysis.
The chemical composition is determined by inductively coupled plasma (ICP) method. The sample is cut into a rectangular bar (5 mm $\times$ 0.7 mm$^2$) along the lattice planes mentioned below by spark erosion after its orientation is determined by backscattering Laue method. Magnetization is measured using a commercial SQUID magnetometer (MPMS, Quantum Design). The electric properties (longitudinal and Hall resistivity) and Seebeck coefficient are measured by a commercial system (PPMS, Quantum Design). The Nernst coefficient is also measured by the same commercial system up to $T\sim\SI{350}{K}$ and the values at a higher temperature are obtained by our homemade system \cite{TomitaHighTemp}.
Single crystals of CoMnSb are employed for the measurements of the electric and thermoelectric transport properties. Throughout this work, electric current $I$ and heat current $Q \propto \nabla T$ is applied along [2$\bar{1}$1] or [11$\bar{1}$], and a magnetic field $H$ is applied along [011] perpendicular to the current direction.
Here, to compare the experimental results with first-principles calculation, the right-handed coordinates system is used, where $x$, $y$, $z$ are taken to be the axes along [2$\bar{1}$1], [11$\bar{1}$], [011], respectively. These directions are displayed on the photo of a single crystal used in our transport measurement (Fig. 1(a)).

\magenta{
First-principles calculations are conducted based on the non-collinear density functional theory \cite{NC1} (DFT) with {\sc OpenMX} code \cite{OpenMX}.
DFT calculations are performed through the exchange-correlation functional within the generalized-gradient approximation and norm-conserving pseudopotentials \cite{PhysRevB.47.6728}. 
The spin-orbit coupling (SOC) is included by using total angular momentum dependent pseudopotentials  \cite{PhysRevB.64.073106}.
The wave functions are expanded by a linear combination of multiple pseudo-atomic orbitals \cite{PhysRevB.67.155108}.
A set of pseudoatomic orbital basis is specified as Mn6.0-$s3p2d2f1$, Co6.0-$s3p2d2f1$, and Sb7.0-$s2p2d2f1$, where the number after each element stands for the radial cutoff in the unit of Bohr and the integer after $s, p, d, f$ indicates the radial multiplicity of each angular momentum component.
The cutoff energy for charge density of 800 Ry and  a $k$-point mesh of $11\times11\times11$ were used.
From the Bloch states obtained in the DFT calculation, a Wannier basis set is constructed by using the {\sc Wannier90} code \cite{Pizzi2020}.
The basis consists of $p$-character orbitals localized at the Sb site, $d$-character orbitals at the Co, and Mn site. Therefore, we consider 208 orbitals for $2\times2\times2$ CoMnSb superstructure including the spin multiplicity.
This set is extracted from 512 bands in the energy window ranging from $-20$ eV to $+30$ eV.
The anomalous Hall conductivity is computed using a $k$-point mesh of $300\times300\times300$~\cite{WU2017}.
}



\begin{figure}[tbp] \centering
  \includegraphics[width=\columnwidth]{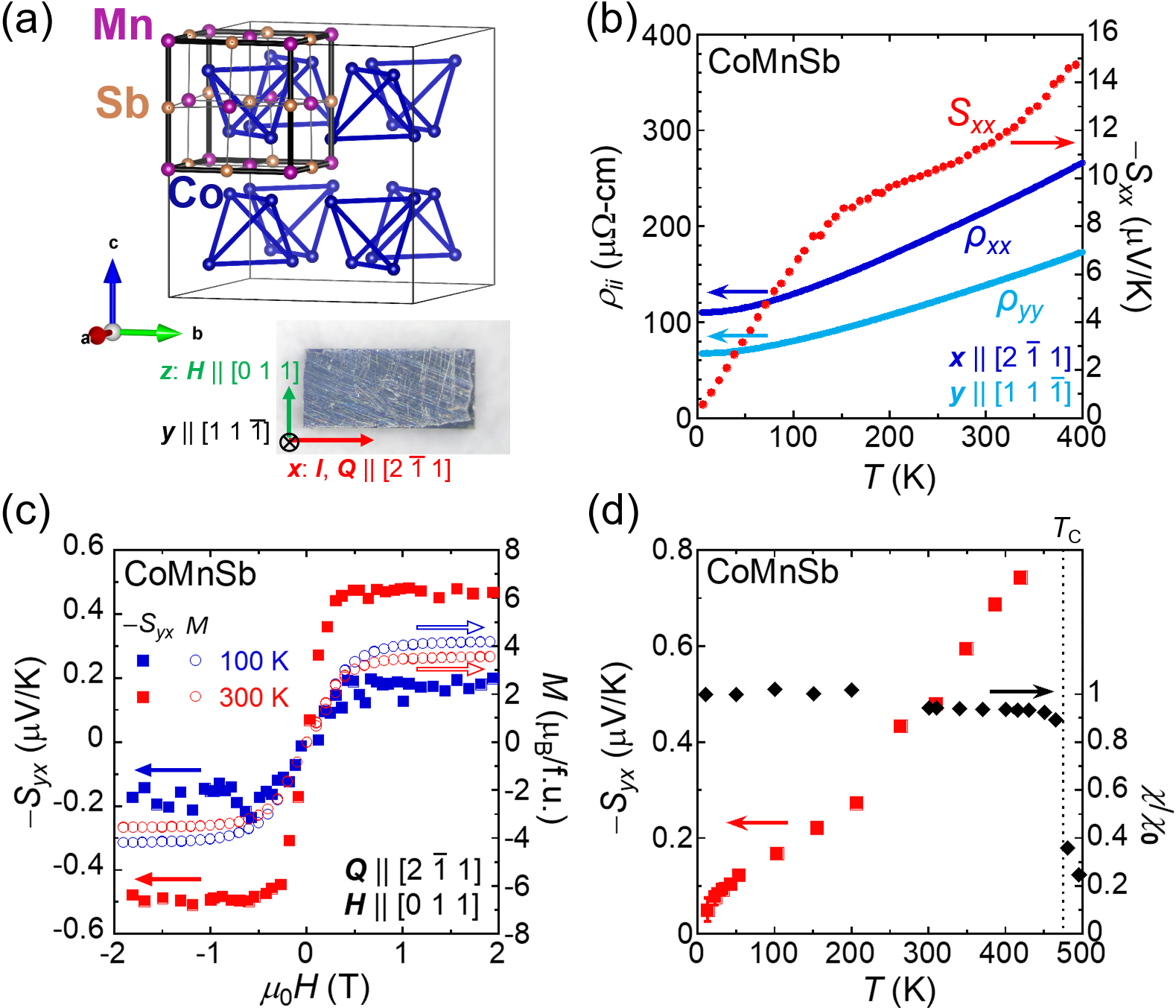}   
  \caption{(a) Superstructure of ferromagnet CoMnSb having 8 unit cells of the half-Heusler one. Mn and Sb atoms are only shown in the single unit cell in this figure for simplicity. Our previous work finds that superstructure CoMnSb has the cubic space group $Fm\bar{3}m$ (No. 225) \cite{Nakamura2020}. The magnetic structure is a collinear ferromagnetism having magnetic moments at both Co and Mn sites \cite{Szytua1972}. Inset: photo of a single crystalline sample of CoMnSb. The electric or heat current is applied along $Q \ || \ [2\bar{1}1]$, and a magnetic field is applied along $H \ ||\  [011]$. (b) Temperature dependence of the longitudinal resistivities $\rho_{ii}$ ($i =x, y$) (left axis) and the Seebeck coefficient $-S_{xx}$ (right axis). (c) Magnetic field dependence of the Nernst coefficient $-S_{yx}$ (left axis) and the magnetization $M$ (right axis) at \SI{100}{K} and \SI{300}{K}. (d) Temperature dependence of the Nernst coefficient $-S_{yx}$ (left axis) and the normalized magnetic susceptibility $\chi/\chi_{0}$ (right axis). $\chi_{0}$ is defined as the magnetic susceptibility at the lowest temperature. The dashed line indicates the ferromagnetic transition temperature  $T_{\mathrm{C}} \sim$ \SI{470}{K} determined in this study.}
\label{fig:1}
\end{figure}

\begin{figure}[tbp] \centering
  \includegraphics[width=\columnwidth]{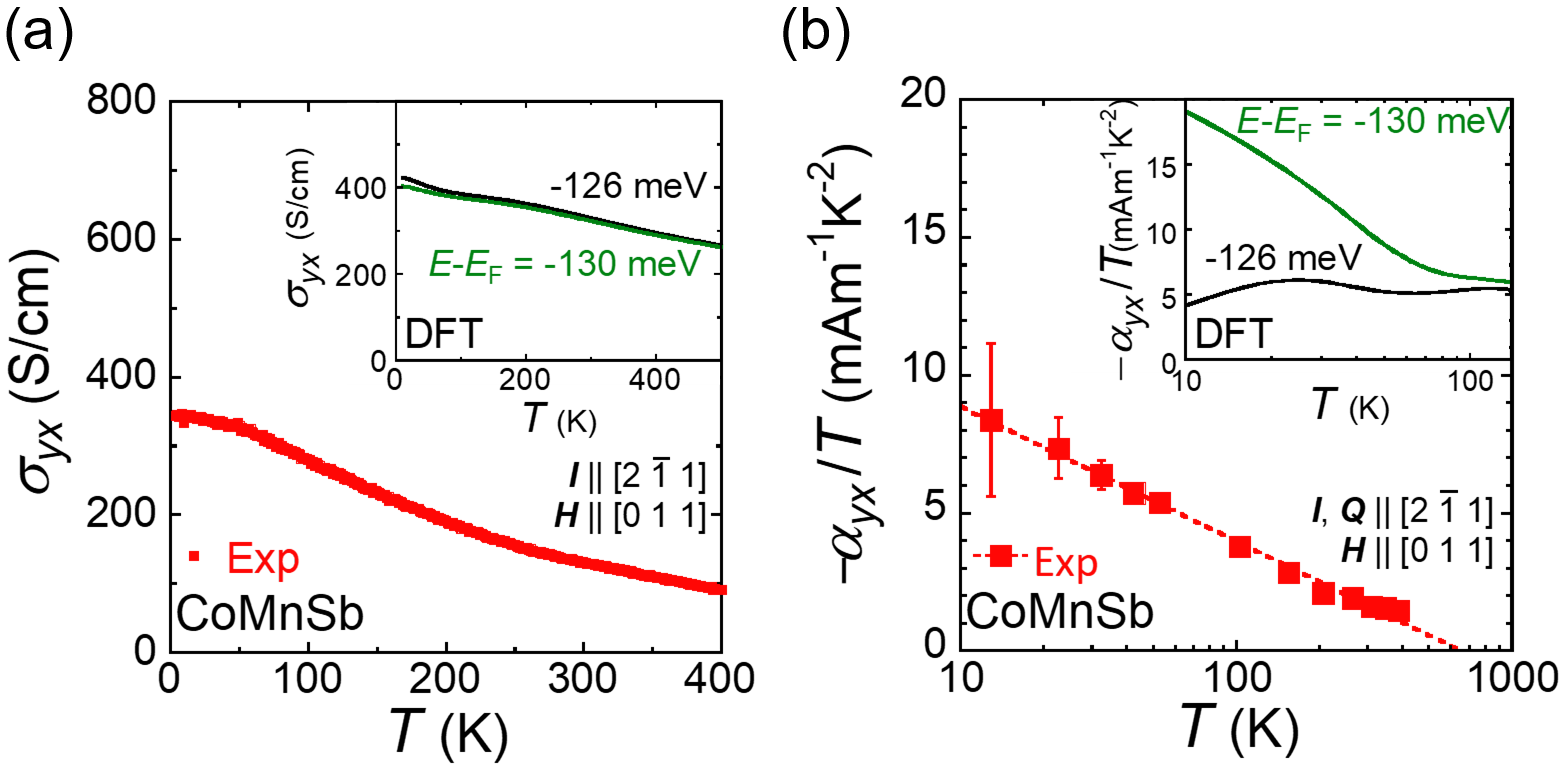}   
  \caption{Temperature $T$ dependence of (a) the Hall conductivity $\sigma_{yx}$, and (b) the transverse thermoelectric conductivity divided by $T$, $\alpha_{yx} / T$ obtained in experiment in comparison with those obtained in theoretical calculations for CoMnSb (insets). Two different chemical potentials $E-E_{\mathrm{F}}=\SI{-126}{meV}$ (black) and $E-E_{\mathrm{F}}=\SI{-130}{meV}$ (green), are selected to display both $\sigma_{yx}(T)$ and $\alpha_{yx}(T) / T$ obtained by the theory in insets, respectively. The error bars are shown if they are larger than the symbol size.}
\label{fig:2}
\end{figure}

First we present the electric transport properties of CoMnSb in Fig. 1(b). The temperature dependence of the longitudinal resistivity $\rho_{ii}(T)$ ($i = x,\ y$) exhibits a metallic behavior with a monotonic increase with temperature in the entire measurement range between 2 and \SI{400}{K}. A slight anisotropy is seen on the $x$-$y$ plane. 

%

Next, we discuss our results on the thermoelectric measurements. 
The thermal gradient of $\nabla_{i} T$ = $-\Delta T / \Delta i$ is applied by flowing a thermal current ${Q_i}$ = $-\nabla_i T$ along the $i$ direction. Here, the temperature difference $\Delta T$ is measured using two thermometers separated by a distance $\Delta i \sim$ 3 mm. Each thermometer is thermally linked to the sample through a strip of copper gold plate. The Nernst voltage $\Delta V_j$ is measured under a magnetic field $H$ along the $j$ direction perpendicular to the thermal gradient $-\nabla_i T$ and the magnetic field. The Nernst coefficient $S_{ji}$ is thus estimated using $-(\nabla_j V) / (\nabla_i T)$ = $-(\Delta V / \Delta T) \cdot (\Delta i / \Delta j)$.
Additionally, from the longitudinal thermoelectric voltage induced by $Q_i$, the Seebeck coefficient is obtained as $S_{ii}$ = $-\nabla_i V / \nabla_i T$ = $- (\Delta V / \Delta T)$.

In the same temperature range as in the resistivity measurements, the Seebeck coefficient is measured at zero magnetic field. The temperature dependence of the Seebeck coefficient $S_{xx}(T)$ is given in the right axis of Fig. 1(b). A notable anomaly is the broad bump observed around \SI{100}{K}. As this temperature is close to $\Theta/5 \  \sim $ \SI{60}{K}, where $\Theta$ is Debye temperature of \SI{275}{K} estimated by the specific heat (not shown here), the bump may be related to the phonon drag effect \cite{Ziman}.
On the other hand, the contribution from the phonon drag in $\alpha_{yx}$ should be negligible since the ANE comes from the intrinsic Berry curvature mechanism as we discuss below. The Nernst angle $\theta_{\rm N}$ ($ = S_{yx} /S_{xx}$) monotonically increases on heating and reaches 5\%, which is comparable in size to the one reported for $L1_{0}$-FePt \cite{Mizuguchi2019}.

Fig. 1(c) presents the magnetic field $\mu_0H$ dependence of both the Nernst coefficient (left axis) and magnetization (right axis) measured at fixed temperatures of 100 and \SI{300}{K}. Due to the difference in the sample shape used for the measurements of the Nernst effect and magnetization, the saturation fields ranges between 0.2 and \SI{1}{T}. For a soft ferromagnet such as CoMnSb, the magnetization and anomalous Nernst effect vanish at zero magnetic field. Therefore, the saturated ones were estimated by taking the average of the values at $|\mu_{0}H|=$1--\SI{2}{T}. The saturated values of the Nernst coefficient $S_{yx}$ significantly increase from \SI{0.2}{\micro V/K} to \SI{0.5}{\micro V/K} by raising temperature from \SI{100}{K} to \SI{300}{K}. On the other hand, no clear change is seen in the size of the saturated magnetization, staying at nearly the same value of 4 $ \mu_{\rm B}/$f.u. under \SI{2}{T}. This magnetization value is consistent with previous experimental reports \cite{Otto1987,Buschow1983} and the DFT calculation made for superstructure CoMnSb \cite{Ksenofontov2006}.

By measuring the field dependence at various temperatures, the temperature dependence of the saturated Nernst coefficient $S_{yx}$ is estimated. Fig. 1(d) shows the temperature dependence of the saturated $S_{yx}$ obtained from 10 to \SI{420}{K}. Notably, $S_{yx} (T)$ shows a significant increase with temperature and reaches \SI{0.7}{\micro V/K} at \SI{420}{K}. This is again in sharp contrast to the temperature dependence of the magnetization obtained in 0.1 T; the magnetization remains nearly constant from \SI{2}{K} up to \SI{470}{K}, where a significant drop is observed on heating across the Curie temperature $T_{\mathrm{C}} \sim $ \SI{470}{K}.

\begin{figure}[tbp] \centering
  \includegraphics[width=\columnwidth]{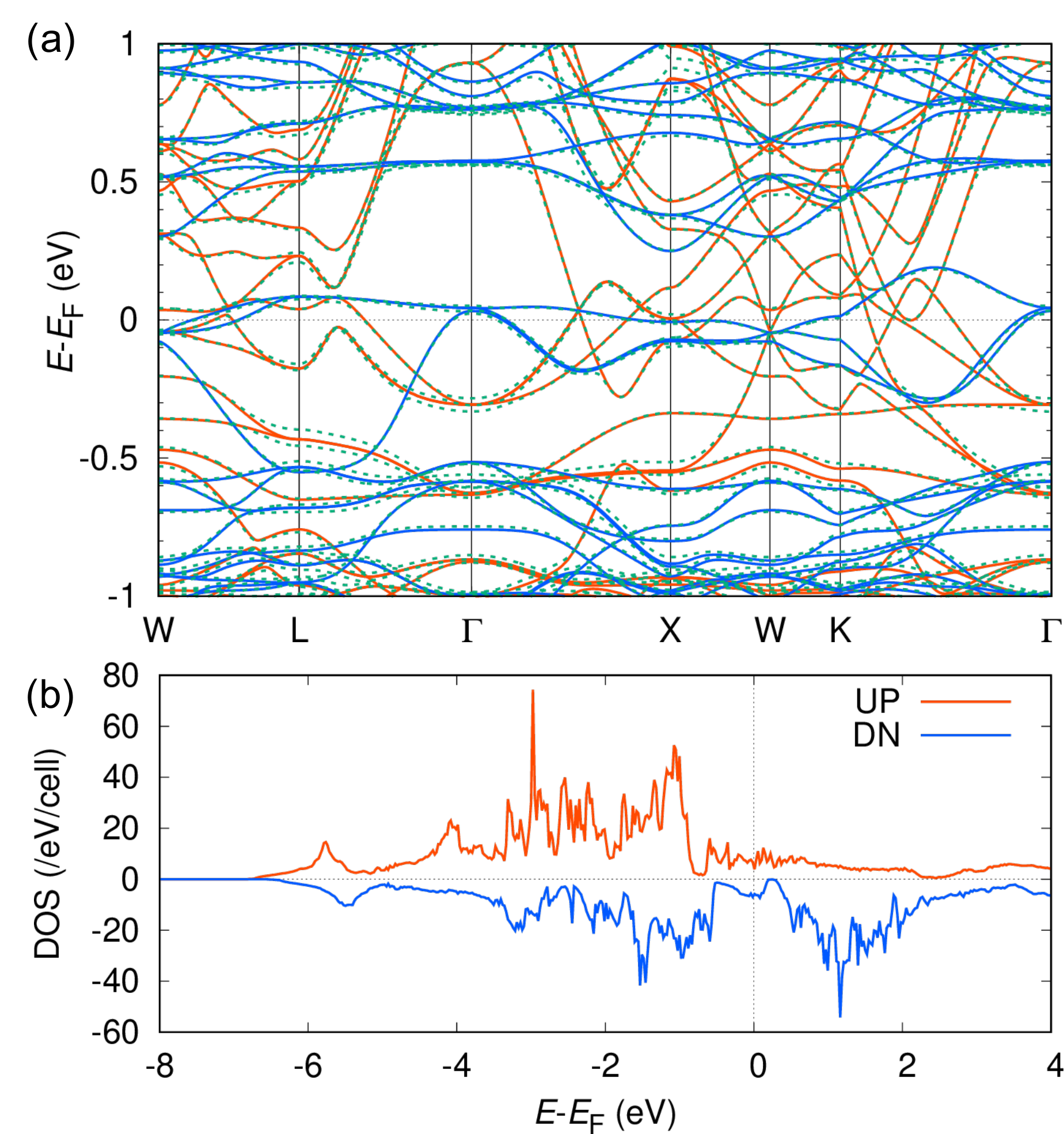}   
  \caption{
  (a) Band structure of CoMnSb. Red and blue solid lines show up and down spin bands, respectively. Green dotted lines show bands including spin-orbit coupling (SOC). (b) Density of states of CoMnSb for up and down spin bands.}
\label{fig:dft}
\end{figure}

To discuss the mechanism of ANE by comparing experiment with theory, let us first estimate the anomalous Hall and transverse thermoelectric conductivities. The Hall conductivity is given by $\sigma_{yx}\approx-\rho_{yx}/\left(\rho_{xx}\rho_{yy}\right)$, where $\rho_{yx}$ is the Hall resistivity and $\rho_{ii}$ is the longitudinal resistivity along $i$-axis. As seen in Fig. 2(a), $\sigma_{yx}$ at \SI{5}{K} for CoMnSb is found as large as \SI{350}{\ohm^{-1}cm^{-1}}, which is even larger than the predicted $\sigma_{yx}$ value in CoMnSb-H \cite{Minami2018}. On heating, $\sigma_{yx}$ decreases monotonically and reaches \SI{100}{\ohm^{-1}cm^{-1}} at \SI{400}{K}.
Using these transport properties, the transverse thermoelectric conductivity can be obtained as $\alpha_{yx}=\sigma_{yy}S_{yx}+\sigma_{yx}S_{xx}$, where $\sigma_{ii}$ is the longitudinal conductivity along $i$-axis ($i=x,y$). Fig. 2(b) shows the temperature dependence of $\alpha_{yx}(T)/T$. The large errors in $\alpha_{yx}$ at \SI{10}{K} and \SI{20}{K} mainly originate from those in $S_{yx}$ because of the small signal size resulting from the Nernst coefficient and temperature difference $\Delta T$.
Notably, $\alpha_{yx}(T)/T$ shows a logarithmic temperature dependence. To clearly see this, our experimental $\alpha_{yx}/T$ is fit to $-\ln T$ using a least-square method (Fig. 2(b) broken line). Over the entire range of the measurements between 10 and \SI{400}{K}, $\alpha_{yx}(T)/T$ exhibits $-\ln T$ behavior within error bars. This is exactly the same behavior seen in the topological magnets such as Co$_2$MnGa \cite{Sakai2018} and Fe$_{3}X$ \cite{Sakai2020} and suggests a similar topological origin.

\begin{figure}[tbp!] \centering
  \includegraphics[width=\columnwidth]{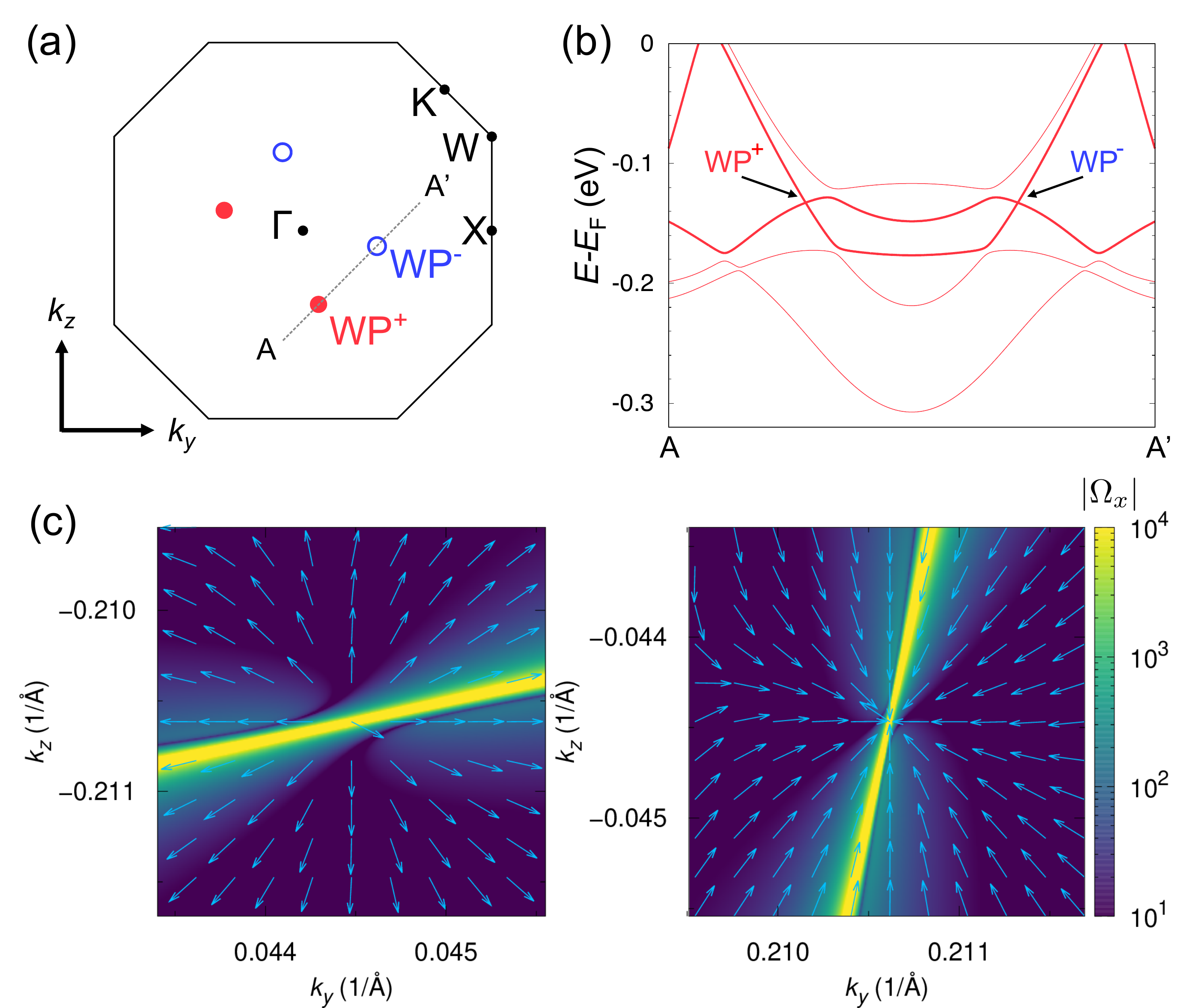}   
  \caption{
  \magenta{
  (a) Distribution of Weyl nodes for CoMnSb in momentum space on the $k_x=0$ plane. The two Weyl node pairs WP$^-$ (open blue circle) and WP$^+$ (closed red circle) are shown. Black solid lines correspond to the boundary of the fcc Brillouin zone.  (b) Band structure along A--A' (the black-dotted line) shown in (a). (c) Berry curvature contribution near the Weyl nodes.
  The contour shows the $x$ component of the Berry curvature ($\Omega_{x}$). The arrows indicate the $y$ and $z$ components of the Berry curvature ($\Omega_{y}$, $\Omega_{z}$). 
  Left and right panel correspond to WP$^+$ and WP$^-$, respectively. These Weyl nodes appear around $E - E_{\rm F} =  -132$ meV.
  }}
\label{fig:weyl}
\end{figure}

To clarify this, the DFT calculations are performed to examine the temperature dependence of $\alpha_{yx}$ in CoMnSb (Fig. \ref{fig:2}(b) inset). 
Here, we estimate its chemical potential dependence by the rigid band approximation.
The anomalous Hall conductivity $\sigma_{yx}$ and the transverse thermoelectric conductivity $\alpha_{yx}$ at finite temperature are calculated through the following formula: $\sigma_{yx}(T,\mu)$ = $e^2/(2\pi)^3\hbar \int d{\bf k} \Omega_{n,z}({\bf k})f_{n,{\bf k}} $,  $\alpha_{yx}(T,\mu) = -1/e \int d\varepsilon \sigma_{yx}|_{T=0} \frac{d f}{d \varepsilon} \frac{\varepsilon - \mu}{T}$, where $e, \hbar, \varepsilon, f, \mu$, and $\Omega_{n,z}$ are the elementary charge with a negative sign, the reduced Planck constant, the band energy, the Fermi-Dirac distribution function with the band index $n$ and the wave-vector ${\bf k}$, the chemical potential and the $z$ component of the Berry curvature, respectively.
By scanning $E-E_{\mathrm{F}}$, $\alpha_{yx} \sim -T\ln T$ behavior is theoretically reproduced at $E-E_{\mathrm{F}}\sim\SI{-130}{meV}$. The sign of this energy deviation is consistent with the experimentally determined sample composition $\mathrm{Co_{0.895}Mn_{0.891}Sb}$, considering the valence electron count (VEC). As can be seen in the inset of \mbox{Fig. 2(b)}, $\alpha_{yx}(T)/T$ exhibits $-\ln T$ behavior down to \SI{10}{K}. 
At the same $E-E_{\mathrm{F}}$, the existence of Weyl node pairs is revealed in the superstructure CoMnSb as discussed below. In contrast, once the $E-E_{\mathrm{F}}$ is shifted slightly from the Weyl nodes such as $E-E_{\mathrm{F}}\sim \SI{-126}{meV}$, the scaling suddenly disappears and gives almost constant $\alpha_{yx}/T$ resulted from the Mott relation.

Finally, let us discuss the existence of the Weyl nodes for CoMnSb.
The electronic structure of CoMnSb around $E - E_{\rm F} \sim \SI{-130}{meV}$ is analyzed based on first-principles calculation. 
\magenta{
Here, the direction of the magnetization is set along the [011] direction.
Fig. \ref{fig:dft}(a) shows the band structure of CoMnSb. The half metallic electronic structure found in  CoMnSb-H is not realized in superstructure CoMnSb \cite{Minami2018} (Fig. \ref{fig:dft}(b)).
The total magnetic moment is $3.99$ $\mu_{\rm B}$/f.u., consistent with experiments including ours and previous one \cite{Ksenofontov2006}.
}
Around at the same $E - E_{\rm F}\sim \SI{-130}{meV}$ where a logarithmic temperature dependence in $\alpha_{yx}/T$  is seen, two pairs of Weyl nodes (four Weyl nodes) are found. 
Fig. \ref{fig:weyl}(a) shows the distribution of the Weyl nodes on the $k_z = 0$ plane. Red solid and blue open circles represent the positive (WP$^+$) and negative (WP$^-$) Weyl nodes, respectively.
Fig. \ref{fig:weyl}(b) shows an enlarged view of the band structure along the dotted line shown in Fig. \ref{fig:weyl}(a).
The pairs are aligned along the \magenta{[011]} direction, which is consistent with the spin configuration used in the calculation.
The monopole-like distribution of the Berry curvature with the opposite chirality distribution are confirmed around each Weyl node, producing large Berry curvature (Fig. \ref{fig:weyl}(c)).
All the above observations indicate that these Weyl nodes should be the origin of the observed $\alpha_{yx}\sim -T\ln T$ behavior.

In this work, we report the transport and magnetization measurements using a single-crystalline sample of the ferromagnet CoMnSb, which is successfully synthesized and has the centrosymmetric superstructure ($Fm\bar{3}m$), distinct from the half-Heusler one having the non-centrosymmetric structure ($F\bar{4}3m$). 
The observed Hall conductivity $\sigma_{yx}$ and Nernst coefficient $S_{yx}$ reached \SI{350}{\ohm^{-1}cm^{-1}} at \SI{5}{K} and \SI{0.7}{\micro V/K} at \SI{420}{K}, respectively and agrees well with the theoretical values. 
Strikingly, the system exhibits a critical $\alpha_{yx}\sim -T\ln T$ behavior deviating from Fermi liquid, which has been seen as a hallmark of the topological ferromagnets such as Co$_2$MnGa \cite{Sakai2018} and Fe$_{3}X$ \cite{Sakai2020}. The logarithmic temperature dependence of $\alpha_{yx}/T$ provides the evidence for the existence of topological texture in momentum space. Our DFT calculations based on the superstructure also reproduce $\alpha_{yx}\sim -T\ln T$ behavior only around $E - E_{\mathrm{F}} \sim$ \SI{-130}{meV}. Moreover, the theory finds that the material hosts four Weyl nodes just around $E - E_{\mathrm{F}} \sim \SI{-130}{meV}$.
Both our theoretical and experimental studies propose that CoMnSb has a ferromagnetic Weyl semimetal state, and the logarithmic \magenta{critical} behavior in a wide temperature renders a key signature for revealing the Weyl semimetal state.

\begin{acknowledgments}
This work is partially supported by CREST (JPMJCR18T3), Japan Science and Technology Agency, by Grants-in-Aid for Scientific Research (19H00650) from the Japanese Society for the Promotion of Science (JSPS). The work for first-principles calculation was supported in part by JSPS Grant-in-Aid for Scientific Research (20K22479). The computations in this research were performed using the Fujitsu PRIMERGY CX400M1/CX2550M5 (Oakbridge-CX) in the Information Technology Center, The University of Tokyo.
The use of the facilities of the Materials Design and Characterization Laboratory at the Institute for Solid State Physics, The University of Tokyo, is gratefully acknowledged. 

\end{acknowledgments}


\bibliographystyle{apsrev4-2}
\bibliography{main}

\end{document}